\documentclass[conference]{IEEEtran}
\IEEEoverridecommandlockouts
\usepackage{cite}
\usepackage{amsmath,amssymb,amsfonts, amsthm}
\usepackage{algorithmic}
\usepackage{graphicx}
\usepackage{textcomp}
\usepackage{xcolor}
\usepackage{hyperref}
\usepackage{algorithmic}
\usepackage{wrapfig}
\usepackage{tabularx}

\usepackage[ruled,linesnumbered]{algorithm2e}
\def\BibTeX{{\rm B\kern-.05em{\sc i\kern-.025em b}\kern-.08em
    T\kern-.1667em\lower.7ex\hbox{E}\kern-.125emX}}

\title{Scheduling Algorithms for Hierarchical Fog Networks}
\author{Amanjot~Kaur and
        Nitin~Auluck}

\begin{document}

\maketitle

\begin{abstract}
Fog computing brings the functionality of the cloud near the edge of the network with the help of fog devices/micro data centers (\emph{mdcs}). Job scheduling in such systems is a complex problem due to the hierarchical and geo-distributed nature of fog devices. We propose two fog scheduling algorithms, named \emph{FiFSA} (Hierarchical \underline{Fi}rst \underline{F}og \underline{S}cheduling \underline{A}lgorithm) and \emph{EFSA} ( Hierarchical \underline{E}lected \underline{F}og \underline{S}cheduling \underline{A}lgorithm). We consider a hierarchical model of fog devices, where the computation power of fog devices present in higher tiers is greater than those present in lower tiers. However, the higher tier fog devices are located at greater physical distance from data generation sources as compared to lower tier fog devices. Jobs with varying granularity and cpu requirements have been considered. In general, jobs with modest cpu requirements are scheduled on lower tier fog devices, and jobs with larger cpu requirements are scheduled on higher tier fog devices or the cloud data center \emph{(cdc)}. The performance of \emph{FiFSA} and \emph{EFSA} has been evaluated using a real life workload trace on various simulated fog hierarchies as well as on a prototype testbed. Employing \emph{FiFSA} offers an average improvement of 27\% and 57.9\% in total completion time and an improvement of 32\% and 61\% in cost as compared to Longest Time First (\emph{LTF}) and cloud-only (\emph{cdc-only}) scheduling algorithms, respectively. Employing \emph{EFSA} offers an average improvement of 48\% and 70\% in total completion time and an improvement of 52\% and 72\% in cost as compared to \emph{LTF} and \emph{cdc-only} respectively.
\end{abstract}


\section{Introduction}
The evolution of IoT devices has lead to the generation of huge amounts of data which needs to be monitored, processed, and analysed {\cite{b1}}. Due to ample computation power and  storage volume, the cloud has become a competent model to execute user applications {\cite{b2}}. The cloud data center is used to process the application data, which is managed by cloud service providers, such as Amazon, Google, or Microsoft Azure. A significant delay in processing the applications has been observed at the cloud data center, owing to its distant geographic location from the users. This delay may be unacceptable for applications with stringent response times, such as smart healthcare {\cite{b3}}. To overcome this delay constraint, system designers are exploring fog computing {\cite{b4, b5}}. 

In general, a cloud only architecture may not be suitable for latency sensitive applications, due to the distance issue highlighted above. Fog computing offers an architecture comprising of a number of fog devices/micro data centres, in close proximity to data generation sources, such as smartphones and sensors. Hence, the generated data can be processed on the fog devices in a timely manner. It has been observed that fog devices have limited computation power and storage capacity \cite{b9}. Hence, they are able to handle jobs with modest cpu requirements, e.g. interactive jobs. Therefore, a flat fog only architecture may not be suitable for processing jobs with less modest execution requirements. This seems to suggest a hierarchical fog-cloud architecture, consisting of multiple fog-tiers and a cloud data center. Accordingly, we consider a hierarchy of fog devices in our architecture \cite{b17}. As an example, consider a 2-tier fog-cloud hierarchy, where tier-1 fog devices have a lower execution capacity than tier-2 fog devices. On the flip side, tier-2 fog devices are located at a greater geographical distance than tier-1 fog devices from the users. Thus, there exists an interesting trade off between execution capacity and propagation delay of tier-1 and tier-2 fog nodes.  The proposed architecture is depicted in Fig.1. Note that this model can be extended to accommodate a larger number of fog tiers, based on application requirements.

\begin{figure}[h]
  \centering
  \includegraphics[width=0.50\textwidth]{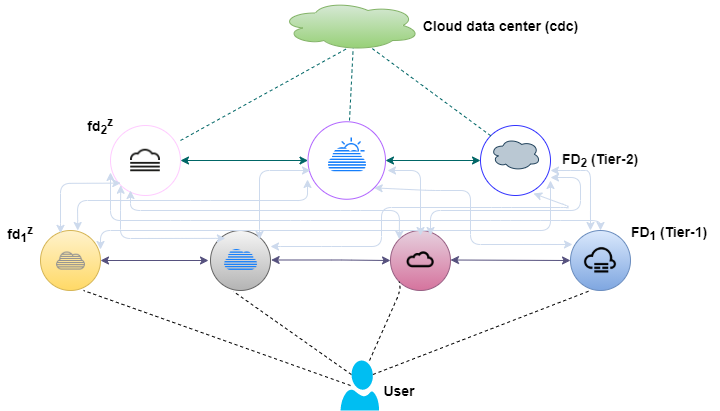}
  \caption{{Fog architecture}}
  \label{fig:Top2}
\end{figure}

As a use case to drive the proposed algorithms, let us discuss a ``smart home''. Typically, a smart home uses various internet connected devices that allows monitoring and supervision. Due to the large number of these devices, such smart homes will generate huge amounts of data per hour. As per the computer business review {\cite{b39}}, it has been reported that UK smart households will generate 26 million GBs of data every week. Some of this data may need real-time responses. Due to network congestion and delayed response times, the cloud data center may not be a suitable option for processing such time-critical data. Handling some of this data at the nearby fog layer may assure real-time response times to users. A smart home scenario can involve a number of  sensors, such as: motion sensors, water sensors, temperature sensors, light sensors, weather sensors etc. A smart home uses the data provided by these sensors to perform various functions such as warning for gas leaks, emergency health services, maintaining the thermostat, managing lights, maintaining the laundry cycle of the washing machine, domestic grocery shopping based on shopping history etc. These jobs will have diverse execution requirements. An example of a job with relatively modest cpu requirements is activating a burglar alarm or a fire detection system. If the alarm isn't generated quickly, then there is a possibility of theft or causalities. It may not be wise to process such data at the cloud data center, due to the significant latency involved. Such jobs should be sent to tier-1 fog devices located in proximity to the user. On the other hand, jobs with medium cpu requirements could be related to maintaining the comfort of the home, such as roller shutter automation, maintaining air-conditioning, switching lights on/off etc. These are examples of jobs with moderate cpu requirements that may be executed at tier-2 fog devices. Lastly, the jobs with high cpu requirements can be executed at the cloud data center. Examples of these jobs could be automatically recording TV programs for subsequent viewing of users. Besides offering reduced response times for job execution, another advantage of fog devices is that they are geographically distributed without a single point of failure, so the service may not be disrupted in the case of device failure.

The main contributions of this paper can be summarised as :

\begin{itemize}
    \item We propose scheduling algorithms for hierarchical fog-cloud architectures with multiple layers of fog devices between the users and the cloud data center. Specifically, we consider flat-tier, 2-tier, 3-tier and 4-tier fog architectures. We formulate our research problem by jointly minimising the completion time as well as the energy consumption on an n-tiered fog-cloud architecture. 
    \item In order to minimise the completion times of the jobs, we propose two hierarchical scheduling algorithms: \emph{FiFSA} (Hierarchical First Fog Scheduling Algorithm), and \emph{EFSA} (Hierarchical Elected Fog Scheduling Algorithm). 
    \item In order to demonstrate the performance of our proposed algorithms, we conduct extensive simulations and develop a working prototype. We run the simulations over a widely accepted simulator - iFogSim, using a real-life Alibaba trace. We observe the system performance by varying a number of parameters such as: job load, delay, and MIPS (Million of Instructions Per Second). We compare our proposed algorithms to another related algorithm for hierarchical fog-cloud architectures - Longest Time First (\emph{LTF}). 
    \item We evaluate and discuss the impact of various fog-cloud hierarchies on the performance of \emph{FiFSA} and \emph{EFSA}.
    
\end{itemize}

The rest of this paper is organized as follows. Section II discusses related work. The model, notation, and problem formulation is described in section III. The proposed algorithms are discussed in section IV. Section V discusses the simulation and experimental results. Finally, section VI concludes the paper and discusses future work.

\section{Related Work}
Cloud computing is a model that provides ``on-demand'' computer services, such as data storage and computation power on a flexible, budget-friendly environment over the internet \cite{b10, b11}.  Cloud computing uses a ``pay as you go'' model, due to which expenses such as hardware costs, manpower costs can be saved by administrators. Even though the cloud offers many advantages, there are multiple challenges: downtime, security, privacy, resource provisioning, power and energy management. On the other hand, IoT has gained widespread popularity with a huge increase in the number of sensors, actuators, and mobile devices connecting billions of things across the world \cite{b12, b13}. The IoT field is approximated to be worth 11 trillion dollars per year by the end of 2025 \cite{b14}. By that time, the deployment of IoT devices is expected to cross 1 trillion. IoT visions a new environment which will impact our day to day life by providing services such as smart healthcare \cite{b15}, smart environmental monitoring\cite{b16}, and so on.  This implies that a large number of applications will be processed and managed by IoT in the future  \cite{b17, b18}. The main requirements in IoT applications are: low latency and fast processing. It may not be possible to achieve low latency by employing cloud computing alone.  The evolution of fog computing has made it possible to meet the demands of the IoT model \cite{b19}. Fog computing extends the services of the cloud closer to the edge of the network, bringing them closer to the users.  The various advantages of fog computing have been discussed in a Cisco white paper \cite{b4}. There are several potential use cases that can benefit from fog computing: smart buildings, autonomous driving, aerial drones etc. The reference architecture of these use cases has been proposed and discussed by the Open Fog Consortium  \cite{b21}. In \cite{b48}, the authors propose an architecture of placing storage and computing nodes close to the users. This architecture mitigates the computation limitation of mobile devices by allowing resource-hungry applications to use cloud-like service for computation. Fog devices are geographically distributed, which helps in maintaining the performance of the system, even in the case of decreased battery power. The modelling and simulation of fog and edge computing environments can be carried out using simulators, such as - iFogSim \cite{b7}. By using iFogSim, one can evaluate the end-to-end latency, achieve Quality of Service (QoS), simulate various resource management strategies, measure power consumption. The simulator follows the sense $\rightarrow$ process $\rightarrow$ actuate model during the simulation of application scenarios. The importance of fog computing in IoT scenarios has been discussed in \cite{b28, b29}. 
 
Multiple papers have explored scheduling on edge computing resources. A run time IoT data placement strategy for latency reduction is proposed in \cite{b24}. In iFogStor, an exact solution is found by using an Integer Linear Programming (ILP) based approach. A heuristic based solution using geographical zoning is proposed in iFogStorZ. This reduces the overall computation time. In \cite{b25}, a survey and a mobility aware scheduling algorithm is proposed. A genetic algorithm is used as a heuristic to provide the possible solution of resource provisioning in the fog-cloud architecture \cite{b26}. In \cite{b27}, the authors provide a comparison of cloud computing and fog computing paradigms. The authors focus on investigating the suitability of fog computing in the IoT environment by providing a network model. In \cite{b30}, the authors use game theory to provide near optimal resource allocation, while increasing the user's quality of experience. The authors propose a hybrid computation offloading approach which minimises the total cost of the system with non orthogonal multiple access \cite{b31}. In \cite{b32}, the authors propose two different matching algorithms to solve the resource allocation problem in two different tiers. In \cite{b33}, a scheduling and resource allocation algorithm is proposed. This algorithm uses Lyapunov optimization to increase the average network throughput of the system. The authors use game theory to formulate a unified multi-tier cost model in a user scheduling problem and prove the occurrence of nash equilibrium \cite{b34}. As the fog nodes have limited computation power, the authors have used a competitive game approach to allocate resources in order to fulfill the user's requests in \cite{b49}. In \cite{b50}, the authors highlight the benefits of fog computing for future IoT industry 4.0 applications. They propose an energy efficient adaptive fog cloud architecture as per application requirements. In \cite{b51}, the authors focus on minimising the latency by using genetic algorithms for IoT applications. However, all these works consider a flat architecture i.e. one layer of fog devices between the edge of the network and the cloud. Hence, they cannot be directly applied to multi-tier fog architectures, which are more complex.

There has been some research in multi-tier hierarchical fog cloud scheduling. In \cite{b43}, Peixoto et al. propose a component based scheduler for a multi-tier fog cloud architecture. They study the effect of shifting the application components to the cloud so as to minimize the impact on network usage. However, they consider only two tiers of cloudlets/fog nodes in their work, and measure the results using  only simulations. We consider a larger number of fog tiers and also report results for a real-life testbed. In \cite{b44}, the authors propose multi-tier real time scheduling algorithms by considering two kinds of priorities -- low and high. We consider non-real time jobs in our work. In \cite{b45}, the authors propose a WALL scheme in which they sort the jobs in descending order based on their workloads (maximum workloads $\rightarrow$ minimum workloads). Next, they sequentially assign each job to a potential cloudlet/fog node that incurs the minimum response delay. Their work is limited to simulations conducted on a smaller number of tiers. In \cite{b42, b44, b46}, the authors propose a Branch and Bound (BnB) algorithm for a multi-tier fog cloud architecture. The proposed algorithm is expensive computationally, and is hard to scale to large applications. This limitation has been addressed by \cite{b46}, where the authors report simulation results for a two-tier fog cloud architecture. In \cite{b47}, the authors propose a latency minimisation algorithm where the tasks are optimally divided and assigned to the nodes on multiple layers. However, their results are restricted only to simulations. On the other hand, we report results for a working prototype testbed.

We observe that various aspects of multi-tier fog computing have been discussed in the above literature. An effective mapping and scheduling of jobs in a multi-tier fog network still faces several challenges, specifically in terms of the cost model. To the best of our knowledge, no work has proposed algorithms for multi-tier fog networks while minimising the execution cost, which comprises of two components: completion time and energy consumption.  Most work multi-tier hierarchical fog networks consists of fewer tiers in their architecture and propose algorithms that may not scale well. Along with this, they have performed either simulations or test-bed for the implementation.  Our research work fills the void by simulating over multi tiers and giving a detailed analysis on the  multi tier performance. We proposed algorithms that can scale well over large systems and validated our results on a hierarchical test-bed.

\section{Problem Formulation}
\begin{table}[h]
\caption{Notation}
\begin{center}
\begin{tabular}{|c|c|}
\hline
${C}$ & cloud data center set\\
${cdc}$ & cloud data center $cdc \in C$\\
${FD}$ & set of all fog devices\\
${FD_1}$ & set of tier-1 fog devices\\
${fd_1^z}$ & ${z^{th}}$ tier-1 fog device $\in FD_1$\\
$FD_2$ & set of all tier-2 fog devices\\
$fd_2^y$ & $y^{th}$ tier-2 fog device $\in FD_2$\\
$FD_n$ & set of all tier-n fog devices\\
$fd_n^x$ & $x^{th}$ tier-n fog device $\in FD_n$\\
$J$ & set of all jobs \\
$N$ & total number of jobs \\ 
$\mu_{fd_t^i}$ & capacity of $i^{th}$ fog device at $t^{th}$ tier\\
$\mu_{cdc}$ & capacity of cloud data center $cdc$\\
$d(j^k, fd_1)$ &  delay between $j^k$ and tier-1 fog device\\
$d(j^k, fd_2)$ &  delay between $j^k$ and tier-2 fog device\\
$d(j^k, fd_n)$ &  delay between  $j^k$ and tier-n fog device\\
$d(j^k, cdc)$ &  delay between  $j^k$ and cloud data center\\
$act(j^k, fd_1)$ & average completion time cost of $j^k$ at tier-1 fog device\\
$act(j^k, fd_2)$ & average completion time cost of  $j^k$ at tier-2 fog device\\
$act(j^k, fd_n)$ & average completion time cost of  $j^k$ at tier-n fog device\\
$aec(j^k, fd_1)$ & average energy consumption cost of $j^k$  at tier-1 fog device\\
$aec(j^k, fd_2)$ & average energy consumption cost of  $j^k$ at tier-2 fog device\\
$aec(j^k, fd_n)$ & average energy consumption cost of  $j^k$ at tier-n fog device\\
$\lambda_i$ & job arrival rate at $i^{th}$ fog device\\
$\Delta_{j^k}$ & cpu requirement of  $j^k$ \\
\hline
 \end{tabular}
 
\end{center}
\end{table}
 
\begin{figure}[h]
  \centering
  \includegraphics[width=0.5\textwidth]{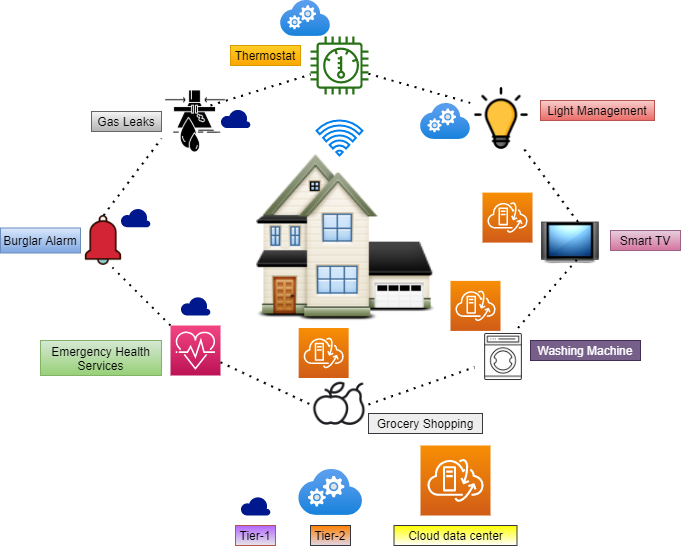}
  \caption{{Fog Services in a Smart Home}}
  \label{fig:SmartHome}
\end{figure}

In this section, we discuss the problem formulation. Table I depicts the notation used in the rest of this paper. We consider a hierarchical fog-cloud architecture comprising of users  ${(u_1, u_2, ....)}$ at the lowest level. The $i^{th}$ user is denoted by $u_i$. The users are connected to \emph{m} tier-1 fog devices given by $FD_1$ = \{$fd_1^1$, $fd_1^2$,......,$fd^{m-1}_1$,$fd^m_1$\}. The tier-1 fog devices are followed by \emph{r} tier-2 fog devices given by $FD_2$ = \{$fd^1_2$, $fd^2_2$,......,$fd^{r-1}_2$,$fd^r_2$\}. The tier-2 fog devices are further connected to \emph{p} tier-n fog devices, such as,  $FD_n$ = \{$fd^1_n$, $fd^2_n$,......,$fd^{p-1}_n$,$fd^p_n$\}. Note that, n can be any tier such as, tier-3, tier-4, and so on. At the topmost level, we have a cloud data center \emph{cdc} $\in$ \emph{C}, where \emph{C} is the set of all cloud data centers. The cloud data center \emph{cdc}  has the maximum execution execution capacity among all the tiers. The execution capacity of fog devices in $FD_1$ is less than  fog devices in $FD_2$, which is further less than $FD_n$.  A similar hierarchical fog architecture has been outlined in \cite{b21}. The
order of execution capacity is: $cdc > FD_n > .... > FD_2 > FD_1$. The units of execution capacity is \emph{MIPS}, i.e., Millions of Instructions per Second. We have used MIPS as the unit of execution capacity, as the iFogSim simulator {\cite{b7}} measures the computational capability of execution devices in MIPS.  iFogSim is a well known simulator that has been used to simulate fog, edge, and cloud networks{\cite{b8}}. A smart home example scenario is shown in Fig.~\ref{fig:SmartHome}. In the considered smart home scenario, the jobs which require a response within milliseconds, such as burglar alarm, gas leaks, etc. can be executed on the fog devices in $FD_1$. The jobs such as maintaining overall temperature of the home, switching lights on/off can be scheduled at fog devices in $FD_2$. Next, the weekly/monthly grocery shopping by tracking the consumption of food of the individuals can be executed at higher tiers  $FD_n$ with the nodes having higher computational capacity, where n = 3. Finally, the entertainment services such as recording some TV series to watch later can use the services of the cloud data center. 

We consider a job set \emph{J}, that can be executed on fog devices in $FD_1$, $FD_2$, ...., $FD_n$, or the \emph{cdc}. We considere a single \emph{cdc} in our work. However, our work can be easily extended to multiple cloud data centers.   The $i^{th}$ fog device  at  tier-1, tier-2, and tier-n  is  given  by $fd^i_1$, $fd^i_2$, and $fd^i_n$ respectively. Likewise, the $k^{th}$ job is given by $j^k$. The delay  between $j^k$ and  $fd^i_1$ is given by $d(j^k,fd^i_1)$. Correspondingly, the  delay  between  $j^k$ and  $fd^i_2$ is given by $d(j^k,fd^i_2)$, and so on. Lastly, $d(j^k, cdc)$ gives the  delay  between  $j^k$ and \emph{cdc}. 
We assume that the fog devices within each tier are homogeneous in nature, i.e., no ``intra-level heterogeneity'' has been considered. This implies that all  fog devices in $FD_1$ have identical execution capacities. This holds true for  fog devices in $FD_2$, and  fog devices in $FD_n$ as well. However, all  tiers, i.e.,   fog devices in $FD_1$,  fog devices in $FD_2$,  fog devices in $FD_n$, and cloud data center \emph{cdc} are heterogeneous with respect to each other. In other words, we consider ``inter-level heterogeneity'' in our model.   Intuitively, jobs with modest $\Delta_{j^k}$ are executed at tier-1 fog devices by default, as they can be executed faster owing to the distance proximity from the user. Generally, the jobs with moderate $\Delta_{j^k}$ are assigned to tier-2, or tier-n fog devices. Finally, the  jobs with large cpu requirements $\Delta_{j^k}$ are assigned to the \emph{cdc}.

Job $j^k$ can be assigned to  fog devices in $FD_1$, $FD_2$, $FD_n$, or the \emph{cdc}, depending upon it's execution cost and/or sufficient spare capacity in the fog device or the cloud. Let $x_{tik}$ be the variable which denotes whether  $j^k$ is assigned to a fog device or to the cloud data center:

\begin{equation}
x_{tik}=\begin{cases}
\text{1}, & \text{if\ $j^k$\ is\ allocated\ to\ $fd_t^i$\ or\ \emph{cdc}}\\
\text{0}, & \text{otherwise}
\end{cases}
\end{equation}

In eq. (1) above, $ fd^i_t$ can be $i^{th}$ fog device present at $t^{th}$ tier ($t=1,2,....,n$).

We need to make sure that each $j^k$ is assigned to a single fog device only.
The following equation ensures this constraint:
\begin{equation}
\sum_{k=1}^{N} x_{tik}=1,  \forall{j^k} \in J
\end{equation}

The begin time \emph{bt} of $j^k$ on a \emph{fd} $\in$ \emph{FD} is denoted by $bt(j^k,fd)$. We assume that all jobs are independent of each other, so no precedence constraints are present among the jobs. Hence, theoretically a job may begin at time 0. The execution time of $j^k$ is represented by $et(j^k,fd)$. The completion time of $j^k$ $\in$ \emph{J} on  \emph{fd} $\in$ $FD$ is designated  by  $ct(j^k,fd)$. The completion time of $j^k$ can be defined as:  
\begin{eqnarray}
   ct(j^k,fd) = bt(j^k,fd) + et(j^k,fd) + d(j^k,fd),\forall j^k \in J
\end{eqnarray}

 The minimum completion time 
 $ct_{min}(j^k)$ of a job can be calculated as:
 \begin{equation}
 ct_{min}(j^k)=ct(j^k, fd^i):ct(j^k, fd^i)<ct(j^k, fd^x)
 \end{equation}
 
In eq. (4) above, $ fd^i, fd^x \in FD$,  $j^k \in J$  and $x \neq i$.

Next, we model the network usage $NU$ of the system.
This quantity is defined as the total data sent and received by various network devices:  fog devices in $FD_1$, $FD_2$, $FD_n$, and the cloud data center \emph{cdc}. Mathematically, $NU$ can be defined as:
\begin{equation}
NU_{sys} =\frac{\sum (dp(j^k) * ad(j^k))}{st} , \forall j^k \in J
\end{equation}

In eq. (5) above, $dp(j^k)$ is the data processed by $j^k$, $ad(j^k)$ is the aggregate delay that occurs during the submission and execution of $j^k$, and $st$ is the time for which the simulation was run.

Let us assume that the maximum amount of jobs that can be processed at  \emph{fd} is  $\mu_i$ and that the job arrival rate at \emph{fd} is $\lambda_i$. Note that, \emph{fd} can be any fog device present at tier-1, tier-2, or tier-n. We follow an M/M/1 queuing system to process user jobs at each fog device \cite{b40}. The average completion time cost $act(j^k,fd)$ for $j^k$ at \emph{fd} can be calculated as:

\begin{equation}
act(j^k,fd)=\frac{1}{\mu_i-x_{tik}\lambda_i}, \forall {fd} \in FD, \forall j^k \in J
\end{equation}

Likewise, we can model \emph{act} of \emph{cdc}  as:
\begin{equation}
act(j^k, cdc) = \frac{1}{\mu_i-x_{tik}\lambda_i}, \forall {cdc} \in C, \forall j^k \in J
\end{equation}

Let the average job execution capacity of \emph{fd} be $acp(fd)$. We can derive the average energy consumption cost $aec(j^k,fd)$ as follows:
\[
aec(j^k,fd) = acp(fd) * \frac{1}{\mu_i-x_{tik}\lambda_i}, \forall {fd} \in FD, \forall j^k \in J \eqno{(8)}
\]

The energy consumption of a fog device is directly proportional to the fog device's computation power/execution capacity.

Similarly,  $aec(j^k,cdc)$ can be defined as follows:
\[
aec(j^k, cdc) = acp(cdc) * \frac{1}{\mu_i-x_{tik}\lambda_i}, \forall {cdc} \in C, \forall j^k \in J \eqno{(9)}
\]

As each fog device has a limited resource capability, this means that the offloaded job's $\Delta_{j^k}$ must not exceed it's  $\mu_i$.
$$     \sum_{k=1}^{N} x_{tik} \lambda_i + \delta_i \leq \mu_i  \eqno{(10)}
$$
Here $\delta_i$ is the remaining spare capacity of \emph{fd} $\in$ \emph{FD}.

We assume that $\Delta_{j^1}$, $\Delta_{j^2}$,..., $\Delta_{j^n}$ be the cpu requirements of the jobs which are being offloaded to fog devices in $FD_1$. The probability that  $fd_1^i$ can run the $j^k$ with $\Delta_k$ is given by  $P(\Delta_{j^k} \leq \mu_{fd_1^i})$. Hence, the probability that all tier-1 fog devices can schedule jobs with different cpu requirements can be defined as follows:
$$
P(\Delta_{j^1} \leq \mu_{fd_1^1}, \Delta_{j^2} \leq \mu_{fd_1^2},......,\Delta_{j^n} \leq \mu_{fd_1^n}) $$
$$= \Pi_{k=1}^{N} P(\Delta_{j^k} \leq \mu_{fd_1^i}) \eqno{(11)}
$$

In case $P(\Delta_{j^k} \geq \mu_{fd_1^i})$, this means tier-1 fog devices are unable to schedule the jobs based on their cpu requirement. In this case, the jobs are sent to the higher tiers for execution. Likewise, we can define the probability for all tier-2 fog devices, and tier-n fog devices. For the sake of brevity, we omit these equations.

The overall cost for all tier-1 fog devices can be written as:
\[
Co_{tier-1} = \sum_{k=1}^{N} \sum_{i=1}^{m} (act(j^k,fd_1^i) + aec(j^k,fd_1^i) ), \forall {fd^i_1} \in FD_1 \eqno{(12)}
\]

The overall cost for all tier-2 fog devices can be written as:
\[
Co_{tier-2} = \sum_{k=1}^{N} \sum_{i=1}^{r}
(act(j^k,fd_2^i) + aec(j^k,fd_2^i)), \forall {fd^i_2} \in FD_2 \eqno{(13)} 
\]

A similar formulation can be carried out for all tier-n fog devices.
\[
Co_{tier-n} = \sum_{k=1}^{N} \sum_{i=1}^{p}
(act(j^k,fd_n^i) + aec(j^k,fd_n^i)), \forall {fd^i_n} \in FD_n  \eqno{(14)} 
\]

Likewise, we model the cost for the cloud:
\[
Co_{cdc} = \sum_{k=1}^{N} 
(act(j^k,cdc) + aec(j^k,cdc)), \forall {cdc} \in C  \eqno{(15)} 
\]

The optimisation problem that we attempt to solve in this work is modelled as follows:
 
\textit{Given a set of jobs $J$, a set of fog devices $FD_1$, $FD_2$, $FD_n$, and a cloud data center \emph{cdc}, with inter-level heterogeneity, schedule the jobs onto fog tiers/cdc by using \emph{FiFSA} \& \emph{EFSA} such that the overall cost is minimised.} Formally,
\begin{equation*}
\begin{aligned}
& \underset{}{\text{minimize}}
& & Co_{sys} = (Co_{tier-1} + Co_{tier-2} + ..... + 
\\
& & & Co_{tier-n} + Co_{cdc}) * x_{tik}  \\
& \text{subject to}
& & \sum_{k=1}^{N} x_{tik}=1,  \forall{j^k} \in J \\ 
& & & \sum_{k=1}^{N} x_{tik} \lambda_i + \delta_i \leq \mu_i  
\end{aligned}
\end{equation*}

\section{Proposed Algorithm}
We now describe the working of the two proposed algorithms for scheduling jobs on multi-tiered fog-cloud architectures : \emph{FiFSA} and \emph{EFSA}. 

\begin{algorithm}
\caption{FiFSA}
\SetKwInOut{KwIn}{Input}
\SetKwInOut{KwOut}{Output}

\KwIn{Queue of jobs $J$ }
\KwOut{\textit{optimalSchedule $J$}}
\begin{algorithmic}[1]
\FOR {k=1 to N} 
 \STATE Arrange jobs in increasing order of $\Delta_{j^k}$ in Queue $Q$
 \STATE flag$(j^k)$ = $0$
  \FOR {i=1 to m} 
  \IF {$(\mu_{fd_1^i} \geq \Delta_{j^k})$}
    \STATE  {Schedule $j^k$ on $fd_1^i$}
      \STATE flag$(j^k) = 1$
     \STATE $break$
    \ELSE
    \STATE {i++}
    \ENDIF
    \ENDFOR
   \FOR {x=1 to r} 
   \IF{$\mu_{fd_2^x}$ $\geq$ $\Delta_{j^k}$}
    \STATE  {Schedule $j^k$ on $fd_2^x$}
     \STATE  flag$(j^k) = 1$
    \STATE  $break$
    \ELSE
    \STATE  {x++}
     \ENDIF
    \ENDFOR
    \IF{ $flag(j^k) = = 0$}
\STATE {Schedule $j^k$ on $cdc$}
\STATE  flag$(j^k)$ = $1$
 \ENDIF
    \ENDFOR
\STATE  Calculate $Co_{sys}$
  
\end{algorithmic}
\end{algorithm}

\begin{algorithm}
\caption{EFSA}
\SetKwInOut{KwIn}{Input}
\SetKwInOut{KwOut}{Output}
\KwIn{Job Set $J$ }
\KwOut{\textit{optimalSchedule }}
\begin{algorithmic}[1]
\STATE {$\forall$  $j^k$ $\in$ $J$ }
\IF {($\mu_{FD_1} \geq \Delta_{j^k})$}
   \STATE {Populate jobs in $Q_1$}
\ELSIF{($\mu_{FD_2} \geq \Delta_{j^k})$}
    \STATE {Populate jobs in $Q_2$}
    \ELSIF{($\mu_{FD_n} \geq \Delta_{j^k}$)}
   \STATE {Populate jobs in $Q_n$}
\ELSE
 \STATE Cloud Scheduler($j^k$)
    \ENDIF
    \IF{($Q_1$,$Q_2$ == Null and $Q_n$ $\neq$ Null)}
    \STATE MinMin($Q_n$, $cl$, $FD_n$)
  \ELSIF{($Q_1$  $\neq$ Null)}
    \STATE MinMin($Q_1$, $cl$, $FD_1$)
  \ELSE
     \STATE MinMin($Q_2$, $cl$, $FD_2$)
     
  \ENDIF
\STATE Calculate $Co_{sys}$
\end{algorithmic}
\end{algorithm}

\begin{algorithm}
\caption{MinMinNode($Q_1$, $cl$, $ct$, $FD_1$)}
\begin{algorithmic}[1]
\STATE $min' \gets \infty$
\FOR{$j \in Q_1$}
\STATE $min \gets \infty$
\FOR{i=1 to m}
\IF{$cl(fd_1^i) + ct(j^k, fd_1^i) < min$}
\STATE $min \gets cl(fd_1^i) + ct(j^k, fd_1^i)$
\STATE $imin \gets i$
\ENDIF
\ENDFOR
\IF{min $<$ min'}
\STATE $ min' \gets cl(fd_1^{imin}) + ct(j^k, fd_1^{imin})$
\STATE $i' \gets imin $
\STATE $j' \gets j^k$
\STATE return(i', j')
\ENDIF
\ENDFOR
\end{algorithmic}
\end{algorithm}

\begin{algorithm}
\caption{MinMin($Q_1$, $cl$, $FD_1$)}
\begin{algorithmic}[1]
\STATE {SchedulingList $SL$ $\gets$ $\Phi$}
\FOR{i=1 to m}
\STATE $cl(fd_1^i) \gets 0$
\FOR{$j \in Q_1$}
\STATE{($i'$, $j'$) $\gets$ $MinMinNode$($Q_1$, $cl$, $ct$, $FD_1$) }
\STATE {Schedule $j'$ on $fd_1^{i'}$}
\STATE {$cl(fd_1^{i'}) \gets cl(fd_1^{i'}) + ct(j', fd_1^{i'} )$}
\STATE{ Remove $j'$ from  $Q_1$, Add $j'$ to $SL$}
\ENDFOR
\ENDFOR
\end{algorithmic}
\end{algorithm}

The steps involved in the working of \emph{FiFSA} are described below:

\begin{enumerate}
    \item In the first step, all the jobs are arranged in increasing order of $\Delta_{j^k}$ in \emph{Q}.
    \item  The $\mu$ of fog devices in $FD_1$ is compared with  $\Delta_{j^k}$. If the current fog device has spare capacity to run  $j^k$, then the job is scheduled on  $fd_1^i$. The value of flag is set as 1.
    \item  Otherwise, the  $\mu$ of  fog devices in $FD_2$ is compared with $\Delta_{j^k}$. If the fog device can run the job, then $j^k$ is scheduled on  $fd_2^i$.
   \item Otherwise, we check the capacity of tier-3 fog devices, and so on till we obtain an appropriate fog tier to run $j^k$. 
    \item Finally, if none of the  $FD_1$, $FD_2$, or  $FD_n$  are capable of scheduling $j^k$, the job is then scheduled on the cloud data center \emph{cdc}.
    \item Calculate  $Co_{sys}$. 
\end{enumerate}

Note that while selecting the fog device to schedule a job, we evaluate all fog devices present at that tier sequentially. Once we identify a fog device which has sufficient spare capacity, we assign the job to the fog device and set the variable flag$(j^k)=1$, and move on the next job. If no fog device at tier-1 is able to schedule the job, we proceed to the tier-2 fog layer, and so on. 

We now discuss the functioning of the proposed scheduling algorithm \emph{EFSA}. Here, we create different queues for each fog-tier according to  $\Delta_j$ of the jobs. The steps involved in the working of \emph{EFSA} are described below:

\begin{enumerate}
    \item The  $\mu$ of fog devices in $FD_1$ is compared with  $\Delta_{j^k}$. If the job's cpu requirement is within the execution capacity of fog devices in $FD_1$,  then the job is populated in $Q_1$. However, if the cpu requirement of job $j^k$ is more than the capacity of that fog-tier, then it is moved to the next tier, else it is placed in the queue of the same tier. 
    \item Otherwise, $\Delta_{j^k}$ is compared with the capacity  $\mu$ of fog node in $FD_2$. If the job can be run with the execution capacity of tier-2 fog devices, then job $j^k$ is added to $Q_2$.
    \item Likewise, we populate the jobs in $Q_n$ by measuring the $\Delta_{j^k}$ with the execution capacity of fog devices present at the tier-n $FD_n$.
    \item The remaining jobs are scheduled on the cloud data center $cdc$.
\end{enumerate}

We use the min-min scheduler to execute the jobs present at  fog devices in $FD_1$, fog devices in $FD_2$, and on  fog devices in $FD_n$. The rationale behind choosing the min-min  scheduler is that it finds the job with smallest completion time and assigns it to the fastest fog device. This fog device is selected by using the procedure MinMinNode($Q_1$, $cl$, $ct$, $FD_1$).

The Min-Min procedure works as follows: 
\begin{enumerate}
    \item For each  $Q_1$, $Q_2$, and $Q_n$ present at each tier $FD_1$, $FD_2$, and $FD_n$ respectively, we use the Min-Min scheduler for the job execution at each tier separately \cite{b41}.
    \item The Min-Min heuristic works in N iterations, where N is the number of jobs which need to be scheduled.
    \item This heuristic consists of two steps, as outlined below:
    \begin{itemize}
        \item In the first step, the heuristic finds the minimum completion time \emph{MCT} of each unassigned job over the set of fog devices. It finds the best fog device i.e. the fog device which can finish the job in the earliest time. The algorithm takes into account the current load $cl(fd_1^i)$ as well as completion time $(ct)$ while finding the best fog device for job scheduling.
        \item In the second step, the heuristic chooses the unassigned job which has the minimum \emph{MCT} among all jobs. This job will be assigned to the best fog device found in the first step.
        \item The scheduled job is removed from the queue, and is not considered in the rest of the iterations.
    \end{itemize}
    \item The above process is repeated until all jobs present in different queues have been scheduled.
    \item Finally, $Co_{sys}$ is calculated.
\end{enumerate}



In \emph{FiFSA}, we first sort the jobs in an increasing order of cpu requirement. We note that the fog devices at low tiers have modest execution capacities. Hence, it may not be advisable to schedule the jobs with high cpu requirements on lower tiers. Once we identify a fog device having sufficient spare capacity, we schedule the job on it. Otherwise, we keep checking fog devices on higher tiers. For the second algorithm, \emph{EFSA} use the MinMin scheduler. MinMin is a fast and simple algorithm that considers smaller jobs first, and assigns them to fast processors, which in turn minimises the completion time. This aligns well with our hierarchical architecture, as the fog devices at lower tiers have lower execution capacity as compared to fog devices present at higher tiers. 

\subsection{Complexity Analysis of \emph{FiFSA} and \emph{EFSA}}

We assume that there are total of $w$ jobs that need to be scheduled, such that $w = w_1 + w_2 $. Here, $w_1$ are the number of jobs that will be scheduled on fog devices and $w_2$ are the number of jobs that will be scheduled on the cloud data centers \emph{cdcs}. Let $t$ be  the total number of fog devices and \emph{cdcs}, such that $t = m + n$. Here, $m$ is the number of fog devices and $n$ is the number of \emph{cdcs}. In \emph{FiFSA}, the jobs are populated in \emph{Q}, for a complexity of $O(w)$. In the next step, the jobs are sorted in increasing order of cpu requirement. The complexity of this stage is $O(w^2)$.  The cpu requirement is checked for each job in the next stage and jobs are assigned to fog devices or \emph{cdcs}. The complexity of this stage is $O(w * t)$. Finally, we calculate the  $Co$ of all jobs. The complexity of this step becomes $O(w + t)$. By adding all these terms, $ O(w * t)$ +  $O(w )$ + $O(w^2)$  +  $O(w + t)$, the overall complexity for \emph{FiFSA} becomes  $\sim$  $O(w^2)$.

In \emph{EFSA}, the cpu requirement is checked for each job in the first step. If the  requirement is met, then the job is added to the corresponding queue. The complexity of this stage is $O(w * t)$. The jobs are either assigned to the \emph{cdc} or to the fog devices, according to the queues. As we are working with a class of assignment problems which are polynomially solvable, this makes the complexity of cloud scheduler $O(n * w_2^2)$ \cite{b22}. We use the min-min scheduler in \emph{EFSA}. The complexity of the min-min scheduler is $O(m * w_1^2)$ \cite{b23}. Finally, we calculate $Co$ for the jobs, for a complexity of $O(m + w_1)$ and $O(n + w_2)$ on fog devices and cdc respectively. On adding all these terms, we get $O(w * t)$ + $O(n * w_2^2)$ +  $O(m * w_1^2)$ +  $O(m + w_1)$ + $O(n + w_2)$. This gives us the complexity expression $\sim$  $O(n * w_2^2)$ +  $O(m * w_1^2)$. 

\section{Simulation Results and Analysis}
We now discuss the results of the simulations that have been carried out to analyse the performance of the proposed algorithms : \emph{FiFSA} and \emph{EFSA}. We consider various simulation scenarios and evaluate the performance for each scenario. The considered sample scenarios align with our fog architecture introduced in Fig 1. The jobs may execute on : tier-1 fog devices, tier-2 fog devices, tier-3 fog devices, tier-4 fog devices, and the cloud data center \emph{cdc}. We have considered several tiers of fog devices to measure the performance of the algorithms on different architectures. The proposed model can be extended to a larger number of tiers, as per the requirements of the application. In \emph{FiFSA}, the jobs execute on the respective tiers as per their  $\Delta$.  If  $\mu$ of a tier-1 fog device is sufficient to meet the requirements of $j^k$, then  $j^k$ will execute on tier-1, otherwise it will be dispatched to tier-2 for execution. Again, $\mu$  of a tier-2 fog device is checked, if it is sufficient, then job can execute on tier-2. If it's not sufficient, then further tiers are checked for job execution. If the job is still unscheduled on all examined fog tiers, it is executed on the \emph{cdc}. We compare the results of four algorithms: \emph{FiFSA}, \emph{EFSA}, \emph{LTF}, and \emph{cdc-only}. In \emph{cdc-only}, the jobs execute only on the cloud data center, without considering the fog devices. The \emph{EFSA} algorithm schedules the job on the fog device by finding the MinMinNode onto various fog tiers. We use MinMinNode and elected fog device interchangeably in the paper. The \emph{LTF} (\underline{L}ongest \underline{T}ime \underline{F}irst) algorithm has been proposed In \cite{b38}. \emph{LTF} assigns the job with the longest execution time to the fog node which executes it in the minimum time i.e fastest node. 

\subsection{Workload}
We now discuss the workload that has been used as an input for our simulation scenarios. In order to make the simulations more meaningful, a real workload trace named the ``Alibaba Cluster Trace'' Program \cite{b35} has been used. The $cluster-trace-v2017$ is the file that consists of cluster information of about 1300 machines. The production cluster consists of both online and batch jobs. The system has been evaluated for a period of 24-hours. This workload can be used to assign the jobs to various machines and cpus, while improving the resource utilisation. It can also be used to explore the trade off in the resource allocation between online services and batch jobs, with a balanced goal of providing better throughput to batch jobs as well as maintaining the satisfactory service quality of the online services. The workload consists of various fields: $timestamp$, $machineID$, $cpu\_utilisation$, $memory\_utilisation$, and $disk\_utilisation$. The fog environment considered for simulation has fog devices in $FD_1$, fog devices in $FD_2$, fog devices in $FD_3$, fog devices in $FD_4$, and a proxy server that is connected to a cloud data center \emph{cdc}. The fog devices in $FD_1$ are connected to various sensors and actuators. We have considered diverse number of fog tiers along with diverse number of fog devices within each tier as per sample architectures. The detailed information about these architectures is provided in Section 5.2. We consider the fog devices within a particular tier to be homogeneous. This means that within a particular tier, the computation capacities of all fog devices are identical. As one moves up in the tiers, the capacity of fog devices increases. On the flip side, moving to higher tiers leads to an increase in the delay between the user and fog devices.

\subsection{Hierarchical Fog-Cloud architecture}

\begin{figure*}[ht]
\begin{minipage}[b]{0.5\linewidth}
\centering
\includegraphics[width=0.8\linewidth]{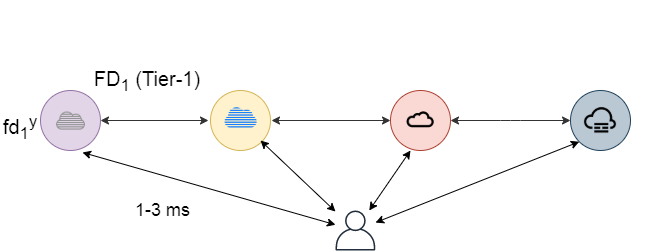}
\caption{Flat hierarchy}
\label{fig:Topology1}
\end{minipage}
\hspace{0.5cm}
\begin{minipage}[b]{0.5\linewidth}
\includegraphics[width=\linewidth]{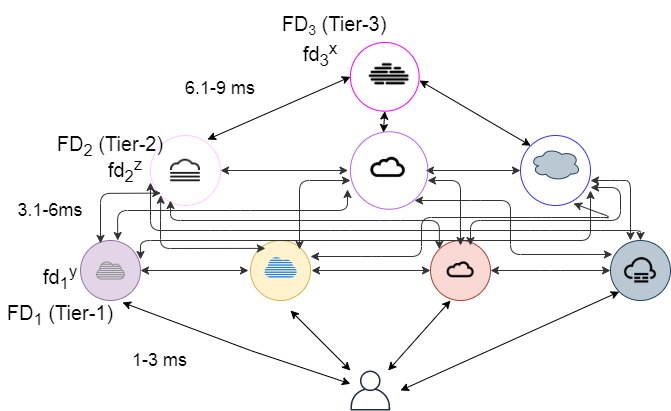}
\caption{3-tier hierarchy}
  \label{fig:Topology4}
\end{minipage}
\end{figure*}

\begin{figure*}[ht]
\centering
\includegraphics[width=0.5\linewidth]{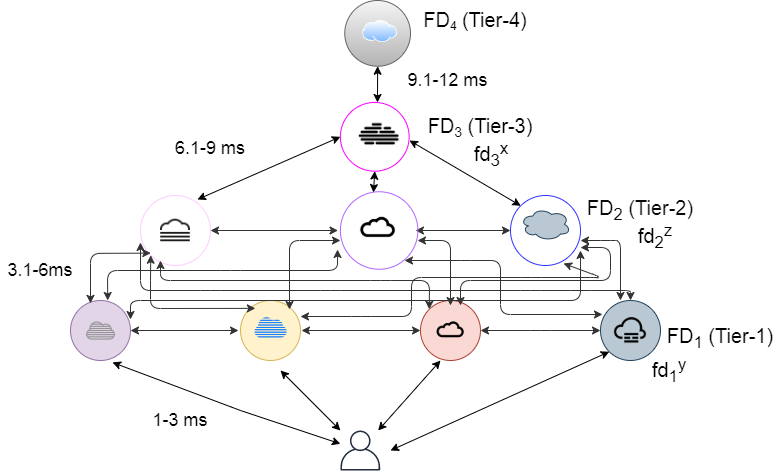}
\caption{4-tier hierarchy}
  \label{fig:Topology6}
\end{figure*}

We consider four types of hierarchical fog cloud architectures in our work. The description of these architectures is as follows:

\begin{itemize}
\item \textbf{Flat hierarchy:} This architecture consists of five fog devices in $FD_1$, followed by the cloud data center \emph{cdc}, as shown in Fig.~\ref{fig:Topology1}. The $d$ from  $u_i$
to $FD_1$ varies from 1 millisecond to 3 milliseconds, and from $u_i$ to  $cdc$ is fixed at 140 milliseconds. The MIPS capacity of $FD_1$ is denoted by $cp(FD_1)$. Likewise, the MIPS capacity of the \emph{cdc} is denoted by $cp(cdc)$. The  $cp(FD_1)$ and $cp(cdc)$ is fixed at 2000 MIPS and 57980 MIPS respectively.

\item \textbf{2-tier hierarchy:} This architecture consists of five fog devices in $FD_1$, followed by five fog devices in $FD_2$, and the cloud data center \emph{cdc} at the top of the hierarchy, as shown in Fig.~\ref{fig:Top2}. The $d$ from  $u_i$
to fog devices in $FD_1$ varies from 1 millisecond to 3 milliseconds, from $u_i$
to fog devices in $FD_2$ varies from 4.1 millisecond to 9 milliseconds and from $u_i$ to $cdc$  is fixed at 140 milliseconds. The $cp(FD_1)$  and  $cp(FD_2)$ is fixed at 2000 MIPS and 3500 MIPS respectively. Likewise, the $cp(cdc)$ is fixed at  57980 MIPS respectively.

\item \textbf{3-tier hierarchy:} In this architecture, we have considered an extension of the 2-tier hierarchy by adding an extra layer of fog devices between fog devices in $FD_2$ and cloud data center \emph{cdc}, as described in Fig.~\ref{fig:Topology4}. We have considered a single fog device at tier-3 of the architecture. The $cp(FD_3)$ is taken as  6500 MIPS. The $d$ between $u_i$ to  fog device in $FD_3$ varies from 10.2 milliseconds to 18 milliseconds.

\item \textbf{4-tier hierarchy:} This architecture is an extension of the 3-tier hierarchy, as shown in Fig.~\ref{fig:Topology6}. We have considered a single fog device in $FD_4$  between  $FD_3$ and \emph{cdc}. The  $cp(FD_4)$ is taken as  8500 MIPS. The $d$ between  $u_i$ to  fog device in $FD_4$ varies from 19.3 milliseconds to 30 milliseconds.
   The $d$ from $u_i$ to $cdc$ is fixed at 140 milliseconds.
\end{itemize}

Note that these are representative values. Based on the user/application specifications, these values can be modified without effecting the functioning of the algorithms. In addition, we have created a representational network for the sake of the simulations. The simulator allows users to vary this as well, based on the requirement.


\subsection{Simulation setup and parameters}

The simulations has been run on an HP workstation with 7.7 GB RAM, Intel core i7-7700 CPU @ 3.60GHz processor, 64-bit Ubuntu OS 18.04 LTS. The proposed algorithms have been implemented on the iFogSim simulator \cite{b7}. Many fog applications have been modeled by researchers using iFogSim \cite{b36, b37}. iFogSim provides a range of functionalities, which makes it a suitable option for simulating various characteristics of fog devices, jobs, and the \emph{cdc}. The simulator consists of a resource management module which manages the resource allocation policies in the fog and cloud environment. In order to implement the proposed algorithms, we created two classes in iFogSim: a \emph{random} class and an \emph{elected} class. The \emph{random} class implements \emph{FiFSA} and the \emph{elected} class implements \emph{EFSA}. The  $\Delta$ and hierarchy of fog devices has been created in these two classes separately. For each hierarchy, we have formed a separate class, such as, \emph{EFSA 4-tier} class, \emph{EFSA 3-tier} class, \emph{EFSA 2-tier} class, \emph{EFSA flat-tier} class,  \emph{FiFSA 4-tier} class, \emph{FiFSA 3-tier} class, \emph{FiFSA 2-tier} class, and \emph{FiFSA flat-tier} class. The parameters such as \emph{d},  \emph{cp(FD)}, \emph{cp(cdc)}, up-link bandwidth, down-link bandwidth and module mapping are described in the classes. 

The parameters that have been used in the simulation scenarios are as follows:

\begin{itemize}
\item \textbf{Job load \emph{(JL)}}: Each job has some MIPS requirement associated with it. The initial MIPS requirement has been taken from the workload trace.  Next, we calculated the average  MIPS  value  for  all  jobs. After the first assignment, the average MIPS value of the jobs is multiplied from 1.5 to 4. Specifically, the MIPS value has been multiplied by 0.5 in each iteration, to obtain various data points. 
\item \textbf{Delay \emph{(d)}:} This parameter defines the range of communication delay between the jobs, $FD_1$, $FD_2$, $FD_3$, $FD_4$, and \emph{cdc}. The parameter \emph{d} from $u_i$ to  $FD_1$ varies from 1 millisecond to 3 milliseconds, \emph{d} between  $FD_1$ to $FD_2$ varies from 3.1 milliseconds to 6 milliseconds,  \emph{d} between $FD_2$ to $FD_3$ varies from 6.1 milliseconds to 9 milliseconds, and \emph{d} between  $FD_3$ to  $FD_4$ varies from 9.1 milliseconds to 12 milliseconds. The parameter \emph{d} between $u_i$ to \emph{cdc} is 140 milliseconds. The initial delay parameters  are taken in the first iteration. After that, the delay is increased by 5 milliseconds at each tier.

 \item \textbf{Network Usage \emph{(NU)}:} This quantity is defined as the total data sent and received by various network devices present at $FD_1$, $FD_2$, $FD_3$, $FD_4$, and the \emph{cdc}.
 
 \item \textbf {Total completion time \emph{(tct)}:} This parameter is defined as the time at which all the submitted jobs complete their execution.
\end{itemize}


\subsection{Effect of fog on system performance}
In this section, we discuss the performance of the proposed scheduling algorithms on multi-tiered fog-cloud architectures. In order to evaluate the performance, we - (i) carry out extensive simulations and (ii) implement a prototype. We discuss the results for flat-tiered, 2-tiered, 3-tiered, and 4-tiered fog architecture, as shown in figures 3, 1, 4 and 5 respectively. This means that the jobs can be processed at: fog devices in $FD_1$, $FD_2$, $FD_3$, $FD_4$, or at the $cdc$.    

\subsubsection{Effect of job load on total completion time} 

\begin{figure}
\centering
\includegraphics[width=0.5\textwidth]{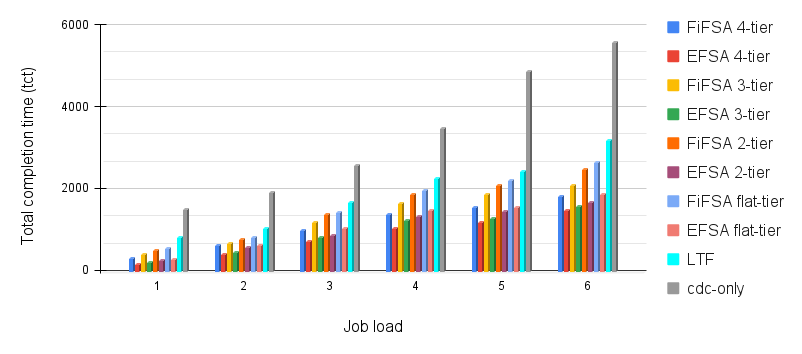}
\caption{Effect of job load on total completion time}
 \label{fig:JlonS}
\end{figure}

The motivation here is to illustrate the advantage of employing fog devices to enhance the capability of the \emph{cdc}. We compare the proposed algorithms with \emph{cdc-only} and \emph{LTF}. In \emph{LTF}, two types of fog devices have been considered : \emph{slow} nodes and \emph{fast} nodes. We consider 5 nodes as slow and 1 node as fast. The average speed of the nodes is calculated in MIPS. The nodes which have a speed less than the average node speed are considered as slow, and the nodes which have a greater speed than the average node speed are considered as fast nodes. We first calculate the average MIPS of the original trace. After this, the average MIPS value is multiplied from 1.5 to 4,  with an interval of 0.5 at each iteration in order to get a range of job loads. The Job Load \emph{(JL)} has been increased, and its effect on the observed \emph{tct} has been recorded for all algorithms. The results are shown in Fig.~\ref{fig:JlonS}. It has been observed that \emph{tct} increases with an increase in the \emph{JL}. In general, increasing the job load adds more computation on the system, which results in increasing the overall \emph{tct}. We observe that the highest \emph{tct} is demonstrated by the \emph{cdc-only} algorithm. The reason for this is that all jobs are allocated to the cloud data center in \emph{cdc-only}, which leads to higher \emph{tct}, owing to the greater physical distance from  $u_i$ to the \emph{cdc}. The \emph{FiFSA} algorithm offers higher \emph{tct} as compared to the \emph{EFSA} algorithm. This can be explained as follows. \emph{FiFSA} assigns the jobs to fog devices on the basis of the \emph{SJF} (Shortest Job First) technique. It maps the shortest job to the first available fog device with sufficient execution capacity. The fog device on which the job is assigned may or may not be best candidate for execution. It may so happen that the job is assigned to a fog device that is farther from the user, therefore there could be an increase in the \emph{tct}. On the other hand, in the \emph{EFSA} algorithm, we calculate the finish times of the jobs on all available fog devices. Next, we assign the job on the elected fog device which ensures the minimum finish time. This results in lower \emph{tct}  for \emph{EFSA}. The \emph{LTF} algorithm assigns jobs with the largest execution time to the fastest node at each level. This makes the waiting time of the short jobs larger, and increases the overall completion time of \emph{LTF}. Hence, we observe that the $tct$ offered by the \emph{LTF} algorithm is more than that offered by both \emph{FiFSA} and \emph{EFSA}, but less than what is offered by \emph{cdc-only}.  

Here, we discuss the effect of job load on \emph{tct} for flat-tiered, 2-tiered, 3-tiered and 4-tiered fog hierarchies. We observe that the best performance is offered by the \emph{4-tier} hierarchy for both proposed algorithms. The reason for this is that the overall computation capacity of the \emph{4-tier} hierarchy is higher than the other hierarchies. This decreases the overall \emph{tct} of the proposed algorithms. Though the fog devices present at $FD_4$ are at greater distance from $u_i$, but sending jobs to tier-4 incurs much less delay as compared to sending jobs to the \emph{cdc}. In addition, the fog devices present at tier-4 have higher computation power than the devices present at lower tiers. The next best performance is offered  by  \emph{3-tier} hierarchy. By removing tier-4 from the architecture, more jobs are sent to the cloud, which increases the \emph{tct}. The decrease in the computation power results in a higher \emph{tct}. The \emph{2-tier} hierarchy performs in between the \emph{3-tier} hierarchy and the \emph{flat-tier} hierarchy. The worst performance is offered by the \emph{flat} hierarchy owing to sending a larger number jobs to the \emph{cdc}, after reaching the computation threshold of tier-1 fog devices. Note that, there isn't much difference in the \emph{tct} of different hierarchies within each algorithm in the starting phase i.e. when job load is less. This happens because  most of the jobs are able to finish execution on the lower tiers initially. However, with the increase in the job load, more jobs are offloaded to the higher fog tiers due to the modest capacity of lower fog devices. Our proposed algorithm optimises the offloading of the jobs by using the \emph{EFSA} algorithm.
We observe that \emph{EFSA} offers a better performance than \emph{FiFSA} for all the hierarchies. This happens as \emph{EFSA} doesn't assign jobs randomly to the fog devices or fog tiers on the basis of \emph{SJF}. It calculates the \emph{MinMinNode} for the job execution. With more number of tiers in \emph{FiFSA}, there are more chances of sending the jobs to the farther fog device once the threshold of the current tier is reached. The \emph{tct} provided by all scheduling hierarchies is as  follows: $EFSA$ 4-tier $< EFSA$ 3-tier $< EFSA$ 2-tier $< EFSA$ flat-tier $<  FiFSA$ 4-tier $< FiFSA$ 3-tier $< FiFSA$ 2-tier $<  FiFSA$ flat-tier $< LTF <$ cdc-only. 

\subsubsection{Effect of job load on network usage}
\begin{figure}
\centering
\includegraphics[width=0.5\textwidth]{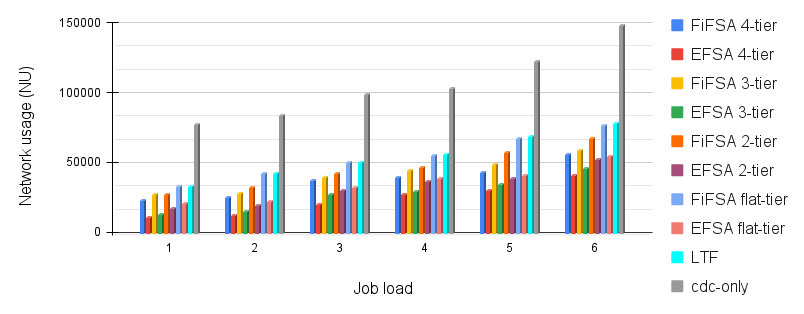}
\caption{Effect of job load on network usage}
 \label{fig:NUonS}
\end{figure}

In this simulation, we observe the effect of job load \emph{(JL)} on overall network usage \emph{($NU$)}. Basically, network usage is the data transferred among various fog devices present at $FD_1$, $FD_2$, $FD_3$, $FD_4$ and the \emph{cdc}. This is calculated by multiplying the transferred data and delay between source and destination and dividing the result by the simulation time. Here, simulation time is the time for which the simulation was run. We observe that when we increase the job load, the data communication of the respective jobs increases as well. The results are shown in Fig.~\ref{fig:NUonS}. It is observed that the network usage is maximum in \emph{cdc-only}. This happens as network usage is directly proportional to the delay incurred, and the data has to travel all the way from the user to the \emph{cdc}. The network usage of the \emph{FiFSA} algorithm is more than that of the \emph{EFSA} algorithm, but less than that of \emph{cdc-only}. It is less than \emph{cdc-only} as \emph{FiFSA} algorithm incorporates fog devices as well, due to which there is lesser propagation delay overall. However, the proposed \emph{EFSA} algorithm doesn't allocate the jobs to fog devices randomly. It finds the elected fog node for the job which finishes the execution in minimum time, for all fog-tiers. Hence, the network usage is minimum for the \emph{EFSA} algorithm. The \emph{LTF} algorithm reports marginally higher network usage than that reported by the \emph{FiFSA} algorithm, as both algorithms use fog devices for job execution. However, the former uses the longest job first technique and the latter uses the shortest job first technique.

Next, we estimate the effect of job load on network usage over flat-tiered, 2-tiered, 3-tiered, and 4-tiered fog hierarchies. The maximum network usage is observed for the flat hierarchy, for both \emph{FiFSA} and \emph{EFSA} algorithms. The reason for this is that with an increase in job load, it becomes hard for a single fog tier to ensure timely job execution. Hence, more jobs are offloaded to the \emph{cdc} for computation, and this increases the delay for these jobs. This leads to an increase in the network usage of the system. The best performance is offered by the \emph{4-tier} hierarchy, as it can accommodate more jobs within the fog tiers, which results in reduced overall transmission delay. The next best performance is offered by the \emph{3-tier} hierarchy. The removal of the fourth fog tier increases the network usage due to the transmission of more jobs from $u_i$ to the \emph{cdc}. With the decrease in the overall computation power offered by the \emph{3-tier} and \emph{2-tier} architectures, the network usage increases in the same order. With large job loads, there is larger reduction in network usage in tiered  hierarchy as compared to the flat hierarchy. The overall network usage of the hierarchies  is as follows:  $EFSA$ 4-tier $< EFSA$ 3-tier $< EFSA$ 2-tier $< EFSA$ flat $< FiFSA$ 4-tier $< FiFSA$ 3-tier $< FiFSA$ 2-tier $<   FiFSA$ flat $< = $LTF $<$ cdc-only.

\subsubsection{Effect of delay on total completion time}

\begin{figure}
\centering
\includegraphics[width=0.5\textwidth]{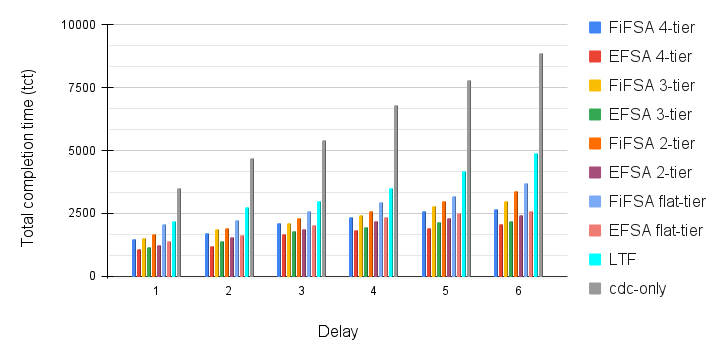}
\caption{Effect of delay on total completion time}
\label{fig:DonS}
\end{figure}

Next, we discuss the effect of propagation delay ($d$, from a job to fog devices or to the \emph{cdc}) on the $tct$. The job load and the fog device count has been kept constant in order to study the effect of delay on the job's \emph{tct}. The delay has been increased by 5 milliseconds at each stage. Intuitively, the \emph{tct} increases with increase in the delay, as the communication becomes a bottleneck. The results of this simulation are shown in Fig.~\ref{fig:DonS}. We observe that by increasing  \emph{d}, the finish times of jobs gets advanced in all algorithms. As mentioned in the previous section, the \emph{EFSA} algorithm outperforms \emph{FiFSA} and \emph{LTF} due to the selection of elected fog node at each tier, which results in  efficient utilisation of fog devices. The \emph{cdc-only} algorithm offers the highest \emph{tct} as fog devices are not employed for job execution. Additionally, the induced delay from  $u_i$ to \emph{cdc} results in increasing the communication time, which in turn increases the \emph{tct}. 

Next, we discuss the effect of delay on the job's \emph{tct} for flat-tiered, 2-tiered, 3-tiered and 4-tiered fog hierarchies.  Similar to previous sections, the \emph{4-tier} hierarchy offers the best results, and the \emph{flat} hierarchy performs the worst, due to the trade-off between computation power and communication delay. This means that \emph{EFSA 4-tier } performs the best among rest of the scheduling strategies i.e. \emph{EFSA 3-tier}, \emph{EFSA 2-tier}, and \emph{EFSA flat-tier}. Likewise, \emph{FiFSA 4-tier } performs the best among rest of the scheduling strategies i.e. \emph{FiFSA 3-tier}, \emph{FiFSA 2-tier}, and \emph{FiFSA flat-tier}.  However, the increased  \emph{d} impacts all four hierarchies, for both \emph{FiFSA} and \emph{EFSA}. This happens as the delay is directly proportional to the job's \emph{tct}. So, the overall \emph{tct} of the jobs gets extended for all four hierarchies, which is visible in the results. The overall  \emph{tct}  is as follows:  $EFSA$ 4-tier $ < EFSA$ 3-tier $< EFSA$ 2-tier $< EFSA$ flat-tier $< FiFSA$ 4-tier $< FiFSA$ 3-tier $< FiFSA$ 2-tier $< FiFSA$ flat-tier $< LTF < cdc-only$. 

\subsubsection{Effect of delay on network usage}

\begin{figure}
\centering
\includegraphics[width=0.53\textwidth]{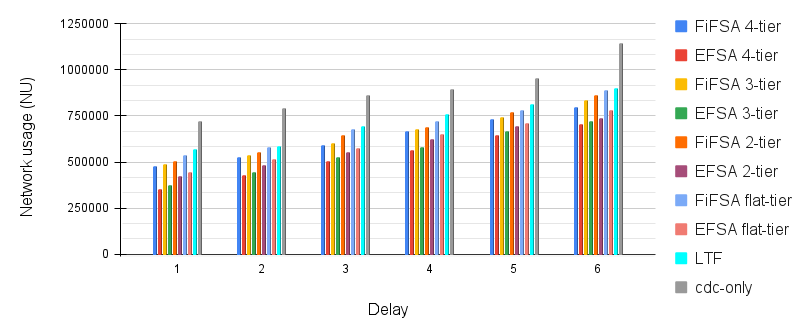}
\caption{Effect of delay on network usage}
\label{fig:Nu1}
\end{figure}

In this simulation, we study the effect of delay on network usage. As network usage is directly proportional to the delay, this means that network usage will increase on increasing the delay. In this simulation, the delay is increased by 5 milliseconds for each point present on the x-axis. We have kept the job load and fog device count constant to measure the effect of delay on network usage. The results are shown in Fig.~\ref{fig:Nu1}. The least network usage is offered by the \emph{EFSA} algorithm. \emph{EFSA} outperforms others as a smaller path needs to be travelled for executing the jobs on their elected fog devices. \emph{FiFSA} and \emph{LTF} perform almost the same in the simulation, as they both use fog devices for executing jobs, but the former chooses fog device randomly from the tiers after applying \emph{SJF}, and the latter assigns the longest job first to run on the fog device that offers minimum execution time. Lastly, in \emph{cdc-only}, the jobs need to travel all way from the user to the cloud, which results in more network usage. 

Next, we observe the effect of delay on network usage over various hierarchies: \emph{EFSA 4-tier}, \emph{EFSA 3-tier}, \emph{EFSA 2-tier}, \emph{EFSA flat-tier},
\emph{FiFSA 4-tier}, \emph{FiFSA 3-tier},
\emph{FiFSA 2-tier}, and \emph{FiFSA flat-tier}. Intuitively, the network usage increases with the increase of delay between fog devices \emph{FD} and \emph{cdc-only}. The least network usage is offered by \emph{EFSA 4-tier} hierarchy. This happens as the algorithm employs four tiers of fog devices, which increases the overall computation power of the system. Along with this, a lesser number of jobs need to travel to the cloud data center, as the jobs can be executed onto the tiers of the architecture which incurs less delay. The \emph{EFSA} algorithm also uses MinMinNode to run the jobs onto several tiers. This decreases the overall network usage of \emph{EFSA 4-tier}. The next best performance is offered by \emph{EFSA 3-tier}, \emph{EFSA 2-tier}, and \emph{EFSA flat-tier} in the same order. This occurs because as we decrease the number of tiers of fog devices, more jobs are sent to the cloud data center for the execution. This increases the network usage of the system. For \emph{FiFSA} algorithm, \emph{FiFSA 4-tier} outperforms all other tiers: \emph{FiFSA 3-tier}, \emph{FiFSA 2-tier}, and \emph{FiFSA flat-tier}. The  \emph{EFSA} outperforms \emph{FiFSA} for all tiers. The reason for this is that the \emph{FiFSA} algorithm can schedule the jobs to the farthest fog device once the capacity of the lower tier fog devices has been exhausted. 

\subsubsection{Effect of change in the number of fog devices on total completion time}

\begin{figure}
\centering
\includegraphics[width=0.5\textwidth]{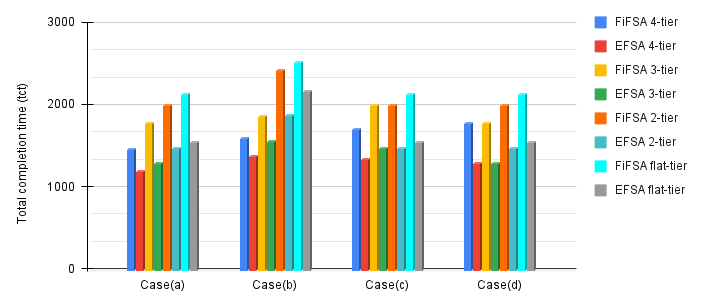}
\caption{Effect of fog device count on total completion time}
\label{fig:PC}
\end{figure}

This section explores the effect of fog processor/device count on the system's \emph{tct}. Specifically, we observe how the algorithms adapt to the change in the number of fog devices.  Initially, we  consider ten fog devices (five at $FD_1$ and five at $FD_2$). The $cp(FD_1)$ and $cp(FD_2)$ has been fixed at 2000 MIPS and 3500 MIPS respectively. We consider a single fog device at $FD_3$ and at $FD_4$. The $cp(FD_3)$ and $cp(FD_4)$ has been fixed at 6500 MIPS and 8500 MIPS respectively. The job load and delay have been kept constant at each stage to focus on the effect of processor count. The results of this simulation are shown in Fig.~\ref{fig:PC}. We consider four cases. In case(a), all the fog devices are considered at all tiers. In case(b), a single fog device is removed, from both $FD_1$ and $FD_2$. In case(c), the lone fog device at $FD_3$  is removed, and in case(d), the fog device at $FD_4$  is removed. For case(a), we can see a similar performance trend of all the hierarchies as discussed in the previous sections. In case(b), as there is a decrease in the fog device count at $FD_1$ and at $FD_2$, we observe an increase in the \emph{tct} for all the hierarchies. However, this increase is more prominent in \emph{FiFSA 2-tier}, \emph{FiFSA flat-tier}, \emph{EFSA 2-tier}, and \emph{EFSA flat-tier}. This occurs due to a decrease in the computation power that leads to advanced \emph{tct}. We observe a marginal increase in the \emph{tct} of \emph{FiFSA 4-tier}, \emph{FiFSA 3-tier}, \emph{EFSA 4-tier}, and \emph{EFSA 3-tier}. The reason for this is that fog devices are present at higher tiers to manage load execution if some fog device become non-functional/ unavailable at lower tiers. In case(c), there is no effect on the \emph{tct} of \emph{FiFSA 2-tier}, \emph{FiFSA flat-tier}, \emph{EFSA 2-tier}, and \emph{FiFSA flat-tier}. These fog device hierarchies do not employ the third tier of fog devices in their architecture. So, their behavior is quite similar to case(a). As the third tier becomes non-functional in  \emph{FiFSA 3-tier} and \emph{EFSA 3-tier}, both of the algorithms start behaving like \emph{FiFSA 2-tier} and \emph{EFSA 2-tier}. We observe an increase in the \emph{tct} for \emph{FiFSA 4-tier} and \emph{EFSA 4-tier}. With a non-functional third tier, more jobs are sent to tier-4 for execution. However, tier-4 is present at a larger distance as compared to tier-3. For case(d), only a 4-tier hierarchy gets affected, as none of the 3-tier, 2-tier, flat-tier hierarchies employ the fourth fog device tier. The algorithms \emph{FiFSA 4-tier} and \emph{EFSA 4-tier} start behaving like their 3-tier equivalents, \emph{FiFSA 3-tier} and \emph{EFSA 3-tier} respectively.

\subsubsection{Effect of MIPS sensitivity on total completion time}
\begin{figure}
\centering
\includegraphics[width=0.5\textwidth]{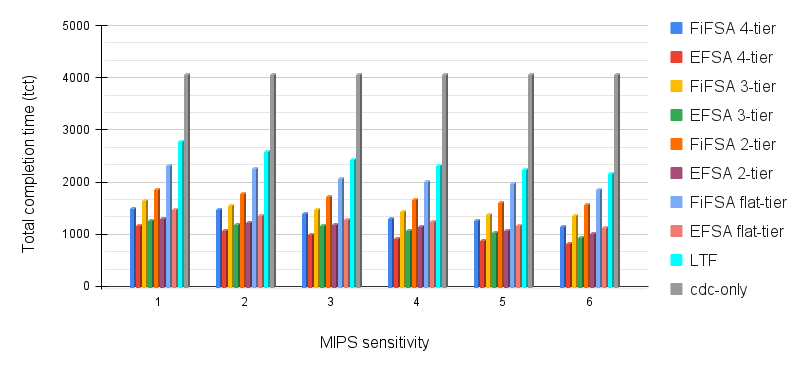}
\caption{Effect of MIPS sensitivity on total completion time }
\label{fig:Mips1}
\end{figure}

In this simulation, we investigate the effect of changing fog device MIPS sensitivity on the system performance. The execution capacity of the \emph{cdc} is the same as in earlier simulations.  The  $cp(FD_1)$ has been varied from 2000 to 3500 MIPS and $cp(FD_2)$ has been varied from 3500 to 5000 MIPS. Likewise, $cp(FD_3)$ has been varied from 6500 to 8000 MIPS and $cp(FD_4)$ has been varied from 8500 to 10000 MIPS.  Essentially, we increase the execution capacity by 300 MIPS at each stage. In the first data point on the x-axis, $cp(FD_1)$, $cp(FD_2)$, $cp(FD_3)$, and $cp(FD_4)$ has been kept at 2000 MIPS, 3500 MIPS, 6500 MIPS, and 8500 MIPS respectively. For the second x-axis data point, these values have been increased by 300 MIPS, and so on. We have kept the job load constant in this simulation. The results of this simulation are shown in Fig.~\ref{fig:Mips1}. As expected, the response time for \emph{cdc-only} is constant for each iteration. On the other hand, the response time for \emph{FiFSA}, \emph{EFSA} and \emph{LTF} decreases with the increase in MIPS capacity. The reason behind this behaviour is that with an increase in the MIPS capacity of fog devices, we are increasing the computation power at each tier. With the increase in the computation power, more jobs can be accommodated by the fog devices, which leads to a decrease in the \emph{tct} of jobs.

Next, we analyse the effect of  MIPS sensitivity on several fog hierarchies. It is evident from the results that with the increase of MIPS values, there is a decrease in the \emph{tct}s. With an increase in computation power, more jobs are able to finish their execution on the fog devices. This decreases the overall job \emph{tct}. The results for this simulation are as follows: $EFSA$ 4-tier $< EFSA$ 3-tier $< EFSA$ 2-tier $< EFSA$ flat-tier $< FiFSA$ 4-tier $< FiFSA$ 3-tier $< FiFSA$ 2-tier $< FiFSA$ flat-tier $<LTF <cdc-only$.

\subsubsection{Effect of job load on cost} 
\begin{figure}
\centering
\includegraphics[width=0.5\textwidth]{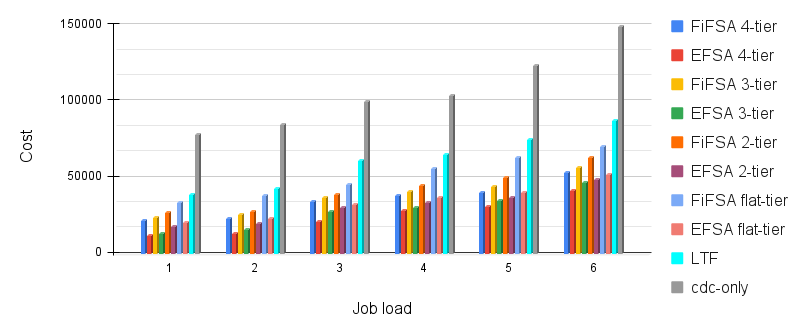}
\caption{Effect of job load on cost}
\label{fig:Cost1}
\end{figure}

In the next simulation, we observe the effect of job load on the system cost $Co_{sys}$. We have taken $cp(FD_1)$ and $cp(FD_2)$ as 2000 MIPS and 3500 MIPS respectively.  Further,  $cp(FD_3)$ and  $cp(FD_4)$ has been picked as 6500 MIPS and 8500 MIPS respectively.  As we can see in Fig.~\ref{fig:Cost1}, the maximum cost is demonstrated by \emph{cdc-only}. The cost is the summation of average  completion time and average energy consumption. Both  completion time and energy consumption are high in \emph{cdc-only}, which leads to a higher  cost. As stated earlier, the average completion time of \emph{FiFSA} is higher than that of \emph{EFSA}. \emph{FiFSA} does not take into account the suitable fog device for executing the job at the respective tiers. This results in higher average completion times, as well as higher average energy consumption, thus making the cost of \emph{FiFSA} more than that of \emph{EFSA}. Additionally, \emph{LTF} offers a higher  cost than \emph{FiFSA} and \emph{EFSA}, as \emph{LTF} takes more time in finishing the jobs. This increases the average completion and average energy consumption of \emph{LTF}. Besides, there is an increase in the cost for all four algorithms \emph{FiFSA}, \emph{EFSA}, \emph{LTF}, and \emph{cdc-only} with the addition of the job load. This happens due to the following - by adding job load on the system, it becomes difficult for fog devices to run the jobs in a timely manner, which in turn increases the overall  cost. 

 With the increase of job load, we observe a gradual increase in the cost for all hierarchies present in \emph{FiFSA} and \emph{EFSA}. The best performance is offered by \emph{EFSA 4-tier} and worst performance is given by \emph{FiFSA flat-tier}. This occurs because \emph{EFSA 4-tier} employs four tiers of fog devices along with the usage of a superior heuristic. On the other hand, \emph{FiFSA flat-tier} has only a single tier of fog devices, and it assigns the jobs on the basis of \emph{SJF} without actually finding the elected fog device. Due to this, cost of \emph{FiFSA flat-tier} is more than \emph{EFSA 4-tier}. We observe a similar trend in tier-wise performance of both the algorithms due to the reasons mentioned in the previous sections.

\subsection{Fog Cloud Testbed}

In order to validate our simulation results, we designed and created a working prototype to evaluate the performance of the two proposed algorithms : \emph{FiFSA} and \emph{EFSA} on a hierarchical fog-cloud architecture. We formed two tiers of fog devices $FD$ followed by the cloud data center \emph{cdc}. We employed two Raspberry Pi 4 Model B devices at  $FD_1$ of the architecture. A desktop with Ubuntu 18.04 operating system has been used as an $FD_2$. Finally, we used a cloud VM instance with 1 vCPU, 1 GiB memory as the \emph{cdc}.  The \emph{cdc} was present in the Asia Pacific (Mumbai) ap-south-1 region. Both raspberry pis are equipped with Broadcom BCM2711, Quad core Cortex-A72 (ARM v8) 64-bit SoC 1.5GHz running Raspbian OS. $FD_2$ is an Intel i7-7700 CPU, 3.60GHz $\times$ 8,  64-bit OS with 7.7GiB RAM. The fog devices were connected through Wifi, which was able to provide service up to a distance of 60 meters. We set the first raspberry pi $fd^1_1$ and second raspberry pi $fd^1_2$ at distances of 2 meters and 10 meters respectively from the WiFi hotspot. The fog device $fd^2_1$ desktop PC was located 20 meters from the WiFi hotspot. We used an IPv4 Internet connection to connect the user to the cloud.

\subsubsection{Total completion time}

\begin{figure}
\centering
\includegraphics[width=0.5\textwidth]{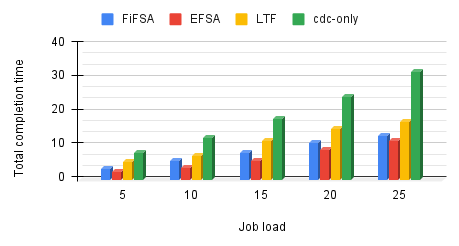}
\caption{Effect of job load on total completion time}
\label{fig:jl1}
\end{figure}

In the first experiment, we observe the impact of job load on \emph{tct} of  \emph{FiFSA}, \emph{EFSA}, \emph{LTF}, and \emph{cdc-only}. Initially, we considered five jobs, and scheduled them using all four scheduling strategies mentioned above. Next, we started increasing the number of jobs (up to 25) in the job set and captured the  \emph{tct}. The results of this experiment are shown in Fig.~\ref{fig:jl1}. In case of \emph{cdc-only}, we observe that the  \emph{tct} increases significantly when we increase the number of jobs in the job set. This happens due to the increase in the overall propagation delay between $u_i$ and \emph{cdc}. On the other hand, the proposed algorithms \emph{FiFSA} and \emph{EFSA} offer lower  \emph{tct}, owing to the incorporation of fog devices for job execution. The  \emph{tct} offered by \emph{LTF} is lower than that offered by \emph{cdc-only}, but worse than the \emph{tct} offered by \emph{FiFSA} and \emph{EFSA}. This is because \emph{LTF} prefers to schedule the long jobs first. This increases the waiting time of the short jobs, leading to higher  \emph{tct}s. 

\subsubsection{Throughput}

\begin{figure}
\centering
\includegraphics[width=0.5\textwidth]{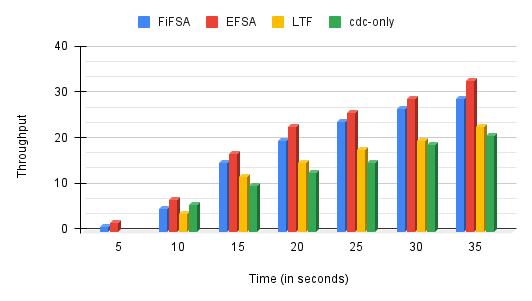}
\caption{Throughput}
\label{fig:th1}
\end{figure}

In the next experiment, we observe the system throughput for all four scheduling strategies. In order to determine the system throughput, we track the number of jobs that finished their execution in the specified amount of time. The results are shown in Fig.~\ref{fig:th1}. As expected, the throughput increases with time for all the four algorithms: \emph{FiFSA}, \emph{EFSA}, \emph{LTF}, and \emph{cdc-only}. This happens because with the increase in the specified time, a larger number of the jobs are able to finish their execution. However, our proposed scheduling algorithms \emph{FiFSA} and \emph{EFSA} offer higher throughput than other compared algorithms due to the employment of fog devices and usage of superior heuristics. The highest throughput is offered by \emph{EFSA}, as it assigns the shortest job to the fog device that can finish the job in the minimum amount of time.

\section{Conclusion}

There is a significant propagation delay involved in executing the jobs on cloud data centers. This delay can be reduced significantly by incorporating fog devices. Application requirements, such as smart transportation, often necessitate the consideration of a hierarchical fog-cloud model, with many layers of fog nodes. In this paper, we propose two algorithms that schedule jobs on such multi-tiered fog networks: \emph{FiFSA} and \emph{EFSA}. We take job characteristics such as cpu requirement into account, while scheduling them on various fog tiers (up to 4 fog tiers) and the cloud data center.  The idea is to allocate the jobs on appropriate fog devices or the cloud data center, while minimising the overall cost. We compared the performance of both algorithms with \emph{cdc-only} and the \emph{LTF} algorithm, for a number of scenarios. The simulation results on a real-life workload and a testbed show that \emph{EFSA} offers the best performance in terms of cost, completion time and network usage. In the future work, we aim to integrate multiple cloud data centers. Additionally, we plan to work on both "inter-level heterogeneity" and "intra-level heterogeneity" in our model.
\bibliography{main}

\begin{thebibliography}{10}

\bibitem{b35}
Alibaba inc. cluster data collected from production clusters in alibaba for
  cluster management research.
\newblock https://github.com/alibaba/clusterdata, 2017.

\bibitem{b51}
Raafat~O Aburukba, Mazin AliKarrar, Taha Landolsi, and Khaled El-Fakih.
\newblock Scheduling internet of things requests to minimize latency in hybrid
  fog-cloud computing.
\newblock {\em Future Generation Computer Systems}, 111:539--551, 2020.

\bibitem{b13}
Ian~F Akyildiz, Max Pierobon, Sasi Balasubramaniam, and Y~Koucheryavy.
\newblock The internet of bio-nano things.
\newblock {\em IEEE Communications Magazine}, 53(3):32--40, 2015.

\bibitem{b11}
Abdulaziz Aljabre.
\newblock Cloud computing for increased business value.
\newblock {\em International Journal of Business and social science}, 3(1),
  2012.

\bibitem{b16}
Bruno And{\`o}, Salvatore Baglio, Antonio Pistorio, Giuseppe~Marco Tina, and
  Cristina Ventura.
\newblock Sentinella: Smart monitoring of photovoltaic systems at panel level.
\newblock {\em IEEE Transactions on Instrumentation and Measurement},
  64(8):2188--2199, 2015.

\bibitem{b23}
Soheil Anousha and Mahmoud Ahmadi.
\newblock An improved min-min task scheduling algorithm in grid computing.
\newblock pages 103--113, 2013.

\bibitem{b2}
Michael Armbrust, Armando Fox, Rean Griffith, Anthony~D Joseph, Randy Katz,
  Andy Konwinski, Gunho Lee, David Patterson, Ariel Rabkin, Ion Stoica, et~al.
\newblock A view of cloud computing.
\newblock {\em Communications of the ACM}, 53(4):50--58, 2010.

\bibitem{b8}
Luiz~F Bittencourt, Javier Diaz-Montes, Rajkumar Buyya, Omer~F Rana, and Manish
  Parashar.
\newblock Mobility-aware application scheduling in fog computing.
\newblock {\em IEEE Cloud Computing}, 4(2):26--35, 2017.

\bibitem{b19}
Flavio Bonomi, Rodolfo Milito, Preethi Natarajan, and Jiang Zhu.
\newblock Fog computing: A platform for internet of things and analytics.
\newblock {\em Big data and internet of things: A roadmap for smart
  environments}, pages 169--186, 2014.

\bibitem{b9}
Rodrigo~N Calheiros, Amir~Vahid Dastjerdi, Harshit Gupta, Soumya~K Ghosh, and
  Rajkumar Buyya.
\newblock Fog computing: Principles, architectures, and applications.
\newblock {\em Internet of things}, pages 61--75, 2016.

\bibitem{b3}
Luca Catarinucci, Danilo De~Donno, Luca Mainetti, Luca Palano, Luigi Patrono,
  Maria~Laura Stefanizzi, and Luciano Tarricone.
\newblock An iot-aware architecture for smart healthcare systems.
\newblock {\em IEEE internet of things journal}, 2(6):515--526, 2015.

\bibitem{b44}
Djabir~Abdeldjalil Chekired, Lyes Khoukhi, and Hussein~T Mouftah.
\newblock Industrial iot data scheduling based on hierarchical fog computing: A
  key for enabling smart factory.
\newblock {\em IEEE Transactions on Industrial Informatics}, 14(10):4590--4602,
  2018.

\bibitem{b4}
Cisco.
\newblock Fog computing the internet of things: Extend the cloud to where the
  things are.
\newblock {\em White Paper}, 2015.

\bibitem{b22}
H~Cui, J~Zhang, C~Cui, and Q~Chen.
\newblock Solving large-scale assignment problems by kuhn-munkres algorithm.
\newblock pages 822--827, 2016.

\bibitem{b36}
Bilal~Khalid Dar, Munam~Ali Shah, Huniya Shahid, and Adnan Naseem.
\newblock Fog computing based automated accident detection and emergency
  response system using android smartphone.
\newblock In {\em 2018 14th International Conference on Emerging Technologies
  (ICET)}, pages 1--6. IEEE, 2018.

\bibitem{b15}
Haluk Demirkan.
\newblock A smart healthcare systems framework.
\newblock {\em It Professional}, 15(5):38--45, 2013.

\bibitem{b46}
Elie El~Haber, Tri~Minh Nguyen, and Chadi Assi.
\newblock Joint optimization of computational cost and devices energy for task
  offloading in multi-tier edge-clouds.
\newblock {\em IEEE Transactions on Communications}, 67(5):3407--3421, 2019.

\bibitem{b45}
Qiang Fan and Nirwan Ansari.
\newblock Workload allocation in hierarchical cloudlet networks.
\newblock {\em IEEE Communications Letters}, 22(4):820--823, 2018.

\bibitem{b21}
OpenFog Consortium Architecture~Working Group et~al.
\newblock Openfog reference architecture for fog computing. 2017.
\newblock {\em URL: https://www. openfogconsortium.
  org/wp-content/uploads/OpenFog\_Reference\_Architecture\_2
  \_09\_17-FINAL.pdf}.

\bibitem{b7}
Harshit Gupta, Amir Vahid~Dastjerdi, Soumya~K Ghosh, and Rajkumar Buyya.
\newblock ifogsim: A toolkit for modeling and simulation of resource management
  techniques in the internet of things, edge and fog computing environments.
\newblock {\em Software: Practice and Experience}, 47(9):1275--1296, 2017.

\bibitem{b12}
Debiao He and Sherali Zeadally.
\newblock An analysis of rfid authentication schemes for internet of things in
  healthcare environment using elliptic curve cryptography.
\newblock {\em Comput {M}ethod {A}ppl {M}}, 2(1):72--83, 2014.

\bibitem{b25}
Bidoura~Ahmad Hridita, Mohammad Irfan, and Md~Shariful Islam.
\newblock Mobility aware task allocation for mobile cloud computing.
\newblock {\em International Journal of Computer Applications}, 137(9):35--41,
  2016.

\bibitem{b41}
O.H. Ibarra and C.E. Kim.
\newblock Heuristic algorithms for scheduling independent tasks on nonidentical
  processors.
\newblock {\em Journal of the ACM (JACM)}, 24(2):280--289, 1977.

\bibitem{b17}
Jiong Jin, Jayavardhana Gubbi, Slaven Marusic, and Marimuthu Palaniswami.
\newblock An information framework for creating a smart city through internet
  of things.
\newblock {\em IEEE Internet of Things journal}, 1(2):112--121, 2014.

\bibitem{b39}
Joao Lima.
\newblock
  https://www.cbronline.com/internet-of-things/10-of-the-biggest-iot-data-generators-4586937/.
\newblock 2015.

\bibitem{b32}
Tingting Liu, Jun Li, BaekGyu Kim, Chung-Wei Lin, Shinichi Shiraishi, Jiang
  Xie, and Zhu Han.
\newblock Distributed file allocation using matching game in mobile fog-caching
  service network.
\newblock In {\em IEEE INFOCOM 2018-IEEE Conference on Computer Communications
  Workshops (INFOCOM WKSHPS)}, pages 499--504. IEEE, 2018.

\bibitem{b31}
Yiming Liu, F~Richard Yu, Xi~Li, Hong Ji, and Victor~CM Leung.
\newblock Hybrid computation offloading in fog and cloud networks with
  non-orthogonal multiple access.
\newblock In {\em IEEE INFOCOM 2018-IEEE Conference on Computer Communications
  Workshops (INFOCOM WKSHPS)}, pages 154--159. IEEE, 2018.

\bibitem{b34}
Zening Liu, Yang Yang, Yu~Chen, Kai Li, Ziqin Li, and Xiliang Luo.
\newblock A multi-tier cost model for effective user scheduling in fog
  computing networks.
\newblock In {\em IEEE INFOCOM 2019-IEEE Conference on Computer Communications
  Workshops (INFOCOM WKSHPS)}, pages 1--6. IEEE, 2019.

\bibitem{b14}
James Manyika, Michael Chui, Peter Bisson, Jonathan Woetzel, Richard Dobbs,
  Jacques Bughin, and Dan Aharon.
\newblock Unlocking the potential of the internet of things.
\newblock {\em McKinsey Global Institute}, 2015.

\bibitem{b50}
Arslan Munir, Prasanna Kansakar, and Samee~U Khan.
\newblock Ifciot: Integrated fog cloud iot: A novel architectural paradigm for
  the future internet of things.
\newblock {\em IEEE Consumer Electronics Magazine}, 6(3):74--82, 2017.

\bibitem{b24}
Mohammed~Islam Naas, Philippe~Raipin Parvedy, Jalil Boukhobza, and Laurent
  Lemarchand.
\newblock ifogstor: an iot data placement strategy for fog infrastructure.
\newblock In {\em 2017 IEEE 1st International Conference on Fog and Edge
  Computing (ICFEC)}, pages 97--104. IEEE, 2017.

\bibitem{b1}
A~Nordrum.
\newblock Popular internet of things forecast of 50 billion devices by 2020 is
  outdated.
\newblock 2016.

\bibitem{b43}
Maycon Peixoto, Thiago Genez, and Luiz~Fernando Bittencourt.
\newblock Hierarchical scheduling mechanisms in multi-level fog computing.
\newblock {\em IEEE Transactions on Services Computing}, 2021.

\bibitem{b18}
Charith Perera, Chi~Harold Liu, Srimal Jayawardena, and Min Chen.
\newblock A survey on internet of things from industrial market perspective.
\newblock {\em IEEE Access}, 2(17):1660--1679, 2014.

\bibitem{b29}
Jurgo Preden, Jaanus Kaugerand, Erki Suurjaak, Sergei Astapov, Leo Motus, and
  Raido Pahtma.
\newblock Data to decision: pushing situational information needs to the edge
  of the network.
\newblock In {\em 2015 IEEE International Multi-Disciplinary Conference on
  Cognitive Methods in Situation Awareness and Decision}, pages 158--164. IEEE,
  2015.

\bibitem{b27}
Subhadeep Sarkar, Subarna Chatterjee, and Sudip Misra.
\newblock Assessment of the suitability of fog computing in the context of
  internet of things.
\newblock {\em IEEE Transactions on Cloud Computing}, 6(1):46--59, 2015.

\bibitem{b48}
Mahadev Satyanarayanan, Paramvir Bahl, Ram{\'o}n Caceres, and Nigel Davies.
\newblock The case for vm-based cloudlets in mobile computing.
\newblock {\em IEEE pervasive Computing}, 8(4):14--23, 2009.

\bibitem{b5}
Mahadev Satyanarayanan, Grace Lewis, Edwin Morris, Soumya Simanta, Jeff Boleng,
  and Kiryong Ha.
\newblock The role of cloudlets in hostile environments.
\newblock {\em IEEE Pervasive Computing}, 12(4):40--49, 2013.

\bibitem{b30}
Hamed Shah-Mansouri and Vincent~WS Wong.
\newblock Hierarchical fog-cloud computing for iot systems: A computation
  offloading game.
\newblock {\em IEEE Internet of Things Journal}, 5(4):3246--3257, 2018.

\bibitem{b49}
Hamed Shah-Mansouri and Vincent~WS Wong.
\newblock Hierarchical fog-cloud computing for iot systems: A computation
  offloading game.
\newblock {\em IEEE Internet of Things Journal}, 5(4):3246--3257, 2018.

\bibitem{b26}
Olena Skarlat, Matteo Nardelli, Stefan Schulte, Michael Borkowski, and Philipp
  Leitner.
\newblock Optimized iot service placement in the fog.
\newblock {\em Service Oriented Computing and Applications}, 11(4):427--443,
  2017.

\bibitem{b42}
Liang Tong, Yong Li, and Wei Gao.
\newblock A hierarchical edge cloud architecture for mobile computing.
\newblock In {\em IEEE INFOCOM 2016-The 35th Annual IEEE International
  Conference on Computer Communications}, pages 1--9. IEEE, 2016.

\bibitem{b40}
Nguyen~H Tran, Cuong~T Do, Shaolei Ren, Zhu Han, and Choong~Seon Hong.
\newblock Incentive mechanisms for economic and emergency demand responses of
  colocation datacenters.
\newblock {\em IEEE Journal on Selected Areas in Communications},
  33(12):2892--2905, 2015.

\bibitem{b37}
Pedro~H Vilela, Joel~JPC Rodrigues, Luciano~R Vilela, Mukhtar~ME Mahmoud, and
  Petar Solic.
\newblock A critical analysis of healthcare applications over fog computing
  infrastructures.
\newblock In {\em 2018 3rd International Conference on Smart and Sustainable
  Technologies (SpliTech)}, pages 1--5. IEEE, 2018.

\bibitem{b10}
William Voorsluys, James Broberg, Rajkumar Buyya, et~al.
\newblock Introduction to cloud computing.
\newblock {\em Cloud computing: Principles and paradigms}, pages 1--44, 2011.

\bibitem{b47}
Pengfei Wang, Zijie Zheng, Boya Di, and Lingyang Song.
\newblock Hetmec: Latency-optimal task assignment and resource allocation for
  heterogeneous multi-layer mobile edge computing.
\newblock {\em IEEE Transactions on Wireless Communications},
  18(10):4942--4956, 2019.

\bibitem{b38}
Hsiang-Yi Wu and Che-Rung Lee.
\newblock Energy efficient scheduling for heterogeneous fog computing
  architectures.
\newblock In {\em 2018 IEEE 42nd Annual Computer Software and Applications
  Conference (COMPSAC)}, volume~1, pages 555--560. IEEE, 2018.

\bibitem{b28}
Marcelo Yannuzzi, Rodolfo Milito, Ren{\'e} Serral-Graci{\`a}, Diego Montero,
  and Mario Nemirovsky.
\newblock Key ingredients in an iot recipe: Fog computing, cloud computing, and
  more fog computing.
\newblock In {\em 2014 IEEE 19th International Workshop on Computer Aided
  Modeling and Design of Communication Links and Networks (CAMAD)}, pages
  325--329. IEEE, 2014.

\bibitem{b33}
Shuang Zhao, Yang Yang, Ziyu Shao, Xiumei Yang, Hua Qian, and Cheng-Xiang Wang.
\newblock Femos: Fog-enabled multitier operations scheduling in dynamic
  wireless networks.
\newblock {\em IEEE Internet of Things Journal}, 5(2):1169--1183, 2018.

\end{thebibliography}

\end{document}